\documentclass[aps, prl, twocolumn]{revtex4-1}
\usepackage{amsmath, amssymb,graphicx, hyperref, color}

\def\aap{Astron. Astrophys.}

\def\mnras{MNRAS}
\def\apj{Astrophys. J.}

\def\apjs{Astrophys. J. Suppl.}

\def\jcap{JCAP}

\begin{document}
\title{The Galactic interstellar medium has a preferred handedness of magnetic misalignment}

\author{Zhiqi Huang $^{1,2}$}
\date{\today}

\affiliation{$^{1}$  School of Physics and Astronomy, Sun Yat-sen University, 2 Daxue Road, Tangjia, Zhuhai, 519082, China \\
  $^{2}$  CSST Science Center for the Guangdong-Hongkong-Macau Greater Bay Area, Sun Yat-sen University, Zhuhai, 519082, China}

\email{huangzhq25@mail.sysu.edu.cn}

\begin{abstract}
  The Planck mission detected a positive correlation between the intensity ($T$) and $B$-mode polarization of the Galactic thermal dust emission. The $TB$ correlation is a parity-odd signal, whose statistical mean vanishes in models with mirror symmetry. Recent work has shown with strong evidence that local handedness of the misalignment between the dust filaments and the sky-projected magnetic field produces $TB$ signals. However, it remains unclear whether the observed global $TB$ signal is caused by statistical fluctuations of magnetic misalignment angles, or whether some parity-violating physics in the interstellar medium sets a preferred misalignment handedness. The present work aims to make a quantitative statement about how confidently the statistical-fluctuation interpretation is ruled out by filament-based simulations of polarized dust emission. We use the publicly available DUSTFILAMENTS code to simulate the dust emission from filaments whose magnetic misalignment angles are symmetrically randomized, and construct the probability density function of $\xi_{p}$, a weighted sum of $TB$ power spectrum. We find that Planck data has a $\gtrsim 10\sigma$ tension with the simulated $\xi_{p}$ distribution. Our results strongly support that the Galactic filament misalignment has a preferred handedness, whose physical origin is yet to be identified.
\end{abstract}

\maketitle

\section{Introduction}

The cosmic microwave background (CMB) is the most powerful cosmological probe to date. Observations of the CMB temperature and polarization anisotropies by the Planck satellite and many other experiments are in good agreement with the standard picture of a Lambda cold dark matter ($\Lambda$CDM) universe that begins with an inflationary epoch~\cite{Planck2018Params, Planck2018Inflation, ACT20, SPT21Params, Polarbear20}. Yet the $B$-mode polarization in CMB, which is regarded as the smoking gun of early-universe inflation, has not been observed~\cite{BK16, BK18, BK21}. The Galactic foreground and, in particular, the thermal emission from interstellar dust grains has become one of the major obstacles to achieving higher-precision measurements of the $B$-mode polarization of CMB~\cite{PlanckXXX, Planck2018FG, BK15}. Thus, the future CMB science crucially depends on the capability of understanding the physics and the statistics of the Galactic thermal dust emission. Since the dust grains tend to align with their short axes parallel to the ambient magnetic field~\cite{Purcell75, DUSTGrain}, the problem of CMB foreground removal is entwined with the study of the Galactic magnetic field, too.

Planck mapped the full sky in nine frequency bands, of which seven are sensitive to polarization. In each polarization-sensitive frequency channel, Planck measures a intensity ($T$) map,  a $Q$-polarization map and a $U$-polarization map. The $Q, U$ maps can be converted to a $E$-mode scalar map and a $B$-mode pseudo-scalar map. The statistics of the maps are often presented in the form of correlation between the $T, E, B$ components, i.e. the $TT, TE, TB, EE, BB, EB$ power spectra.  The $T,\ E$ components  are invariant under a parity transformation, whereas the $B$ component changes sign in the mirror world. Thus, for models that preserve mirror symmetry, the ensemble averages of parity-odd $TB$ and $EB$ power spectra are expected to vanish. A detection of $TB$ or $EB$ correlation beyond statistical fluctuations, either in the Galactic foreground or in the CMB, would provide valuable information of parity-breaking physics beyond the standard picture~\cite{Bracco19, Eskilt22, Greco22, Diego22, Komatsu22}.

The channel at 353\,GHz, the highest polarization-sensitive frequency, is the most sensitive to the Galactic polarized dust emission. The Planck 353\,GHz polarized sky map exhibits a few non-trivial properties, such as a significantly non-unity $EE/BB \sim 2$, a positive $TE$ correlation, and a weakly positive $TB$ correlation~\cite{PlanckXXX, Planck2018FG, Weiland20}. All these features except for the parity-odd $TB$ signal can be qualitatively explained by the state-of-the-art magneto-hydrodynamic (MHD) simulations~\cite{MHD1, MHD2, MHD3} and by phenomenologically modeling the magnetized filamentary structure of the interstellar medium (ISM)~\cite{PlanckEBAsym, DUSTFILAMENTS}. In particular, the publicly available filament-based code DUSTFILAMENTS~\cite{DUSTFILAMENTS} is able to reproduce the $EE$, $BB$, and $TE$ power spectra, as well as some non-Gaussian features (Minkowski functionals) of the Planck 353\,GHz sky map. The DUSTFILAMENTS code is also shown to be in good  agreement with MHD simulations.

It has been demonstrated with strong evidence that the parity-odd $TB$ signal in Planck high-frequency maps is driven by misalignment between dust filaments and the sky-projected magnetic field~\cite{Huffenberger20,Clark21}. However, a globally positive $TB$ correlation over a wide range of scales ($\ell\sim 40$-$600$), as seen in the Planck data, has only been reproduced in phenomenological simulations with artificial input of significantly skewed distribution of the magnetic misalignment angles. For simple models with symmetric distribution of the magnetic misalignment angles, the $TB$ correlation arising from statistical fluctuations is typically much smaller than the observed one~\cite{Huffenberger20, DUSTFILAMENTS}. Thus, the observed $TB$ signal can be regarded as a hint about a global preference of misalignment handedness. Quantification of the statistical significance of such a hint, the topic of the present work, is valuable for the exploration of parity-breaking mechanism in the ISM beyond the current understanding of the Galactic physics~\cite{Planck2018FG, Bracco19}.

The obvious strategy is to simulate dust emission maps with null hypothesis, that is, symmetric distribution of magnetic misalignment angles, and compare the $TB$ correlation of the Planck data and that of the simulated maps. The problem of the direct comparison method is that even phenomenological filament-based simulations of Galactic dust emission are computationally expensive, and the dust $TB$ power spectrum may not obey a multi-variable Gaussian distribution. Straightforward detection of a  $\sim 5\sigma$ anomaly in the $TB$ power spectrum would require at least $\sim$ millions of simulated dust maps, which, taking the DUSTFILAMENTS code for example, costs $\gtrsim 10^9$ CPU hours.

The present work aims to find a more computationally economic approach to estimate the statistical significance of the global preference of misalignment handedness. The idea is to compress the information into a simple integrated quantity. Although the dust intensity and polarization maps are highly non-Gaussian, an integrated quantity contains sums of many adjacent modes, and therefore approximately obeys a Gaussian distribution, whose parameters can be estimated with an affordably small set of simulations.


\section{Data and Softwares}

Following Section 3.3.1 of Ref.~\cite{PlanckXXX}, we produce a series of nested ``large region'' (LR) masks: LR72, LR63, LR53, LR42, LR33, and LR24, where the postfix in LRmn represents the fraction (mn\%) of the sky that is unmasked. To avoid possible bias from noise correlations, we compute the correlation between two different half mission maps of Planck data release~3~\cite{Planck2018HFI}. We have confirmed that all the results presented in this paper do not vary much if we use, say, even-odd splitting of the data~\cite{Planck2018HFI} or the SRoll maps with reduced large-scale systematic effects~\cite{SRoll2}.

We use DUSTFILAMENTS, a code recently developed by Ref.~\cite{DUSTFILAMENTS}, to simulate dust maps. DUSTFILAMENTS is a three dimensional model composed of $\sim 10^8$ filaments that are imperfectly aligned with the magnetic field. Although the filament-only recipe is likely to be an oversimplification of the complex morphology of the ISM~\cite{Konstantinou22}, DUSTFILAMENTS is able to reproduce the main features of the Planck 353\,GHz maps. In particular, it generates frequency decorrelation and non-Gaussian features on small scales that are both in agreement with Planck data~\cite{DUSTFILAMENTS}. By default, we use the software settings suggested in Table~1 of Ref.~\cite{DUSTFILAMENTS}, which assumes a symmetric distribution of the magnetic misalignment angles, the null hypothesis that we want to test.

All the maps are processed with the standard Healpix software~\cite{Healpix}. Angular power spectra of masked maps are computed with NaMaster~\cite{NaMaster}, whose $\ell$-bins are taken to be uniformly spacing with $\Delta \ell = 20$. Cross power spectra between two different maps are always symmetrized. For instance, the $TB$ power spectrum $C_{\ell}^{TB}$ actually refers to $\frac{C_\ell^{TB}+C_{\ell}^{BT}}{2}$ in NaMaster convention.

\section{Stacking the Dust Maps}

Before carrying out a detailed statistical description, we would like to qualitatively demonstrate the anomalous $TB$ signal in the Planck data by stacking the polarization maps. The polarization stacking approach, which we sketch below, can be found in details in Refs.~\cite{WMAP7, Planck2016IandS}. 

With flat-sky approximation, the spin-2 $(Q, U)$ component around a central pixel can be written as
\begin{equation}
  Q \pm i U \approx \left(\partial_x\pm i\partial_y \right)^2\nabla^{-2}(E\pm i B).
\end{equation}
The $\nabla^{-2}$ operator is defined in Fourier space as a multiplier $-\frac{1}{k^2}$. The local Cartesian coordinates $x, y$ are given by
\begin{equation}
  x = 2\sin\frac{\theta}{2}\cos\phi, \  y = 2\sin\frac{\theta}{2}\sin\phi, \label{eq:xydef}
\end{equation}
where $\theta$ is the angular distance from the central pixel, and $\phi$ is the angle between the radial vector from the central pixel and the longitudinal vector (pointing to the south pole). We use $\omega \equiv 2\sin\frac{\theta}{2}$ instead of $\theta$ to approximate the radial distance, because the transformation given by Eq.~\eqref{eq:xydef} preserves the pixel area from spherical sky to flat sky\footnote{For the $4^\circ\times 4^\circ$ stacking in the present work, the difference between $\theta$ and $2\sin\frac{\theta}{2}$ is actually unimportant.}. The $(Q_r, U_r)$ components are obtained by rotating the $(Q, U)$ components from the local $(\vec{e}_x, \vec{e}_y)$ Cartesian basis to the radial and tangential polar basis,
\begin{eqnarray}
  Q_r &=& Q\cos(2\phi) + U \sin(2\phi), \\
  U_r &=& U\cos(2\phi) - Q \sin(2\phi).
\end{eqnarray}
The correlations between $Q_r, U_r$ and the intensity at the central pixel, $T(0)$, are given by
\begin{eqnarray}
  \langle Q_r(\omega)T(0)\rangle &=& -\int\frac{\ell d\ell}{2\pi} C_{\ell}^{TE}\mathrm{J}_2(\ell\omega), \label{eq:Qr} \\
  \langle U_r(\omega)T(0)\rangle &=& -\int\frac{\ell d\ell}{2\pi} C_{\ell}^{TB}\mathrm{J}_2(\ell\omega), \label{eq:Ur} 
\end{eqnarray}
respectively. Here $C_\ell^{TE}$ and $C_\ell^{TB}$ are the $TE$ and $TB$ power spectra, and $\mathrm{J}_2$ is the Bessel function of the first kind of order $2$. The extra minus signs on the right-hand side of Eqs.~(\ref{eq:Qr}-\ref{eq:Ur}) arise from the Bessel integral $\frac{1}{2\pi}\int_0^{2\pi}e^{i(\ell\omega\cos\phi - 2\phi)}d\phi = - J_2(\ell\omega)$.

According to Eqs.~(\ref{eq:Qr}-\ref{eq:Ur}), the patterns of stacked $Q_r$, $U_r$ around pixels with a skewed distribution of $T(0)$ explore the $TE$, $TB$ correlations, respectively. Because $TB$ correlation has a much lower signal to noise ratio than $TE$, the conventional approach used in the literature, that stacking polarization maps around intensity peaks, does not yield a recognizable pattern. The signal to noise ratio can be enhanced if we stack $U_r$ around many more hot pixels. To focus on the scales where the anomalous $TB$ signal is found, we filter both the intensity map and the polarization maps with a $100\lesssim \ell \le 600$ band-pass filter, defined by
\begin{equation}
  F(\ell) = \left\{
    \begin{array}{ll}
      0, & \text{ if } \ell\le 90 \text{ or } \ell > 600; \\
      1, & \text{ if } 110 \le \ell \le 600; \\
      \sin^2\frac{(\ell - 90)\pi}{40}, & \text{ if } 90< \ell < 110.
    \end{array}
  \right. \label{eq:bandpass}
\end{equation}

\begin{figure*}
  \centering
    \includegraphics[height=0.32\textwidth]{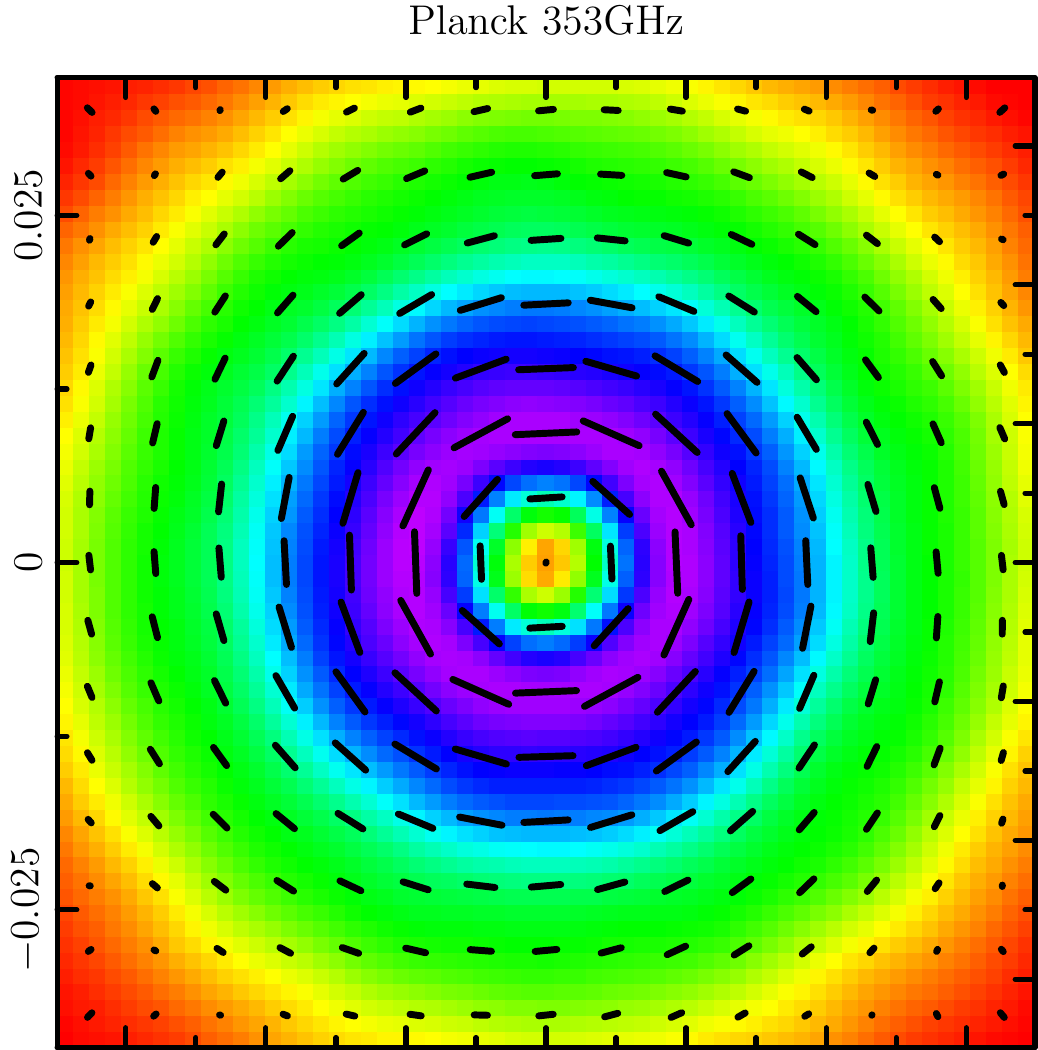}  \includegraphics[height=0.32\textwidth]{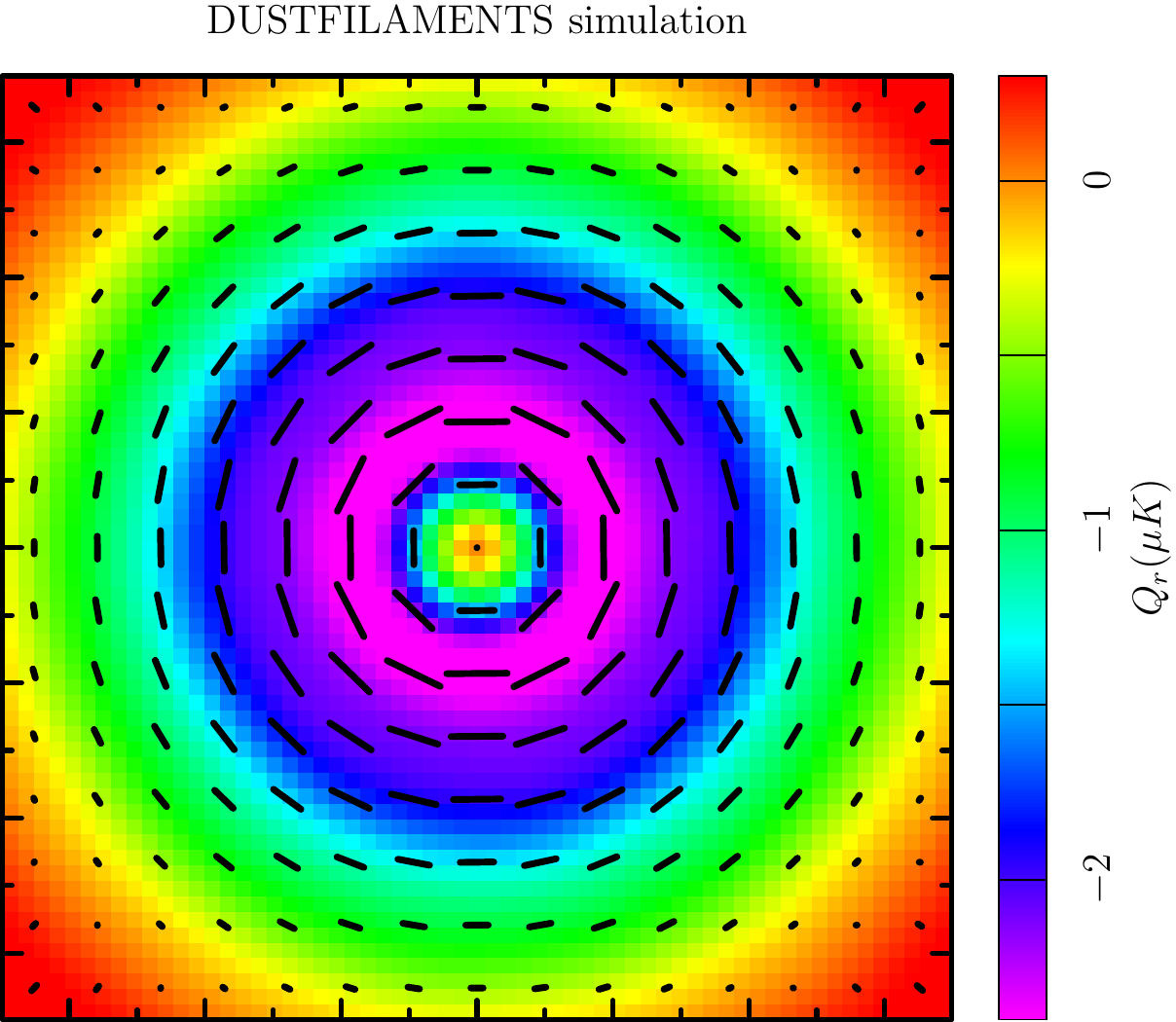} \\
  \includegraphics[height=0.32\textwidth]{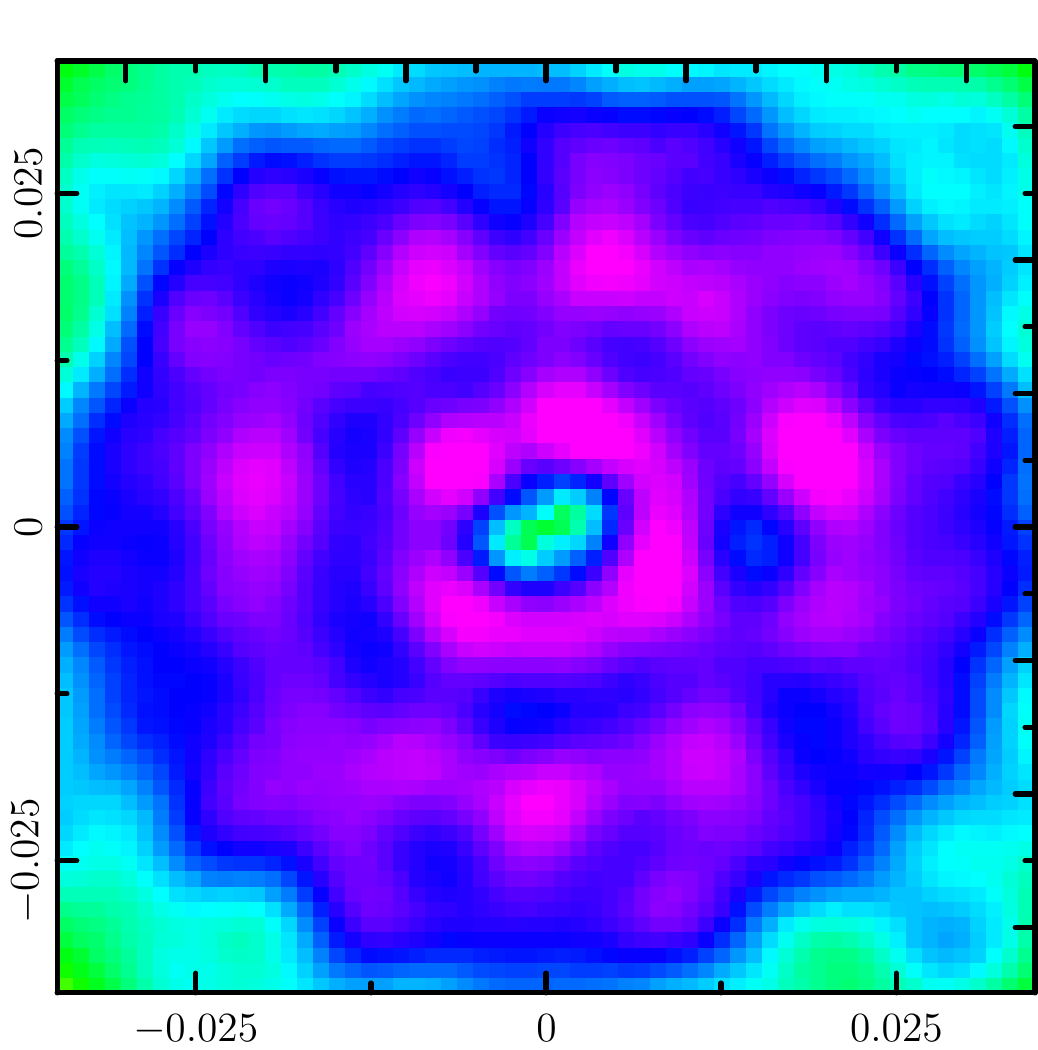}  \includegraphics[height=0.32\textwidth]{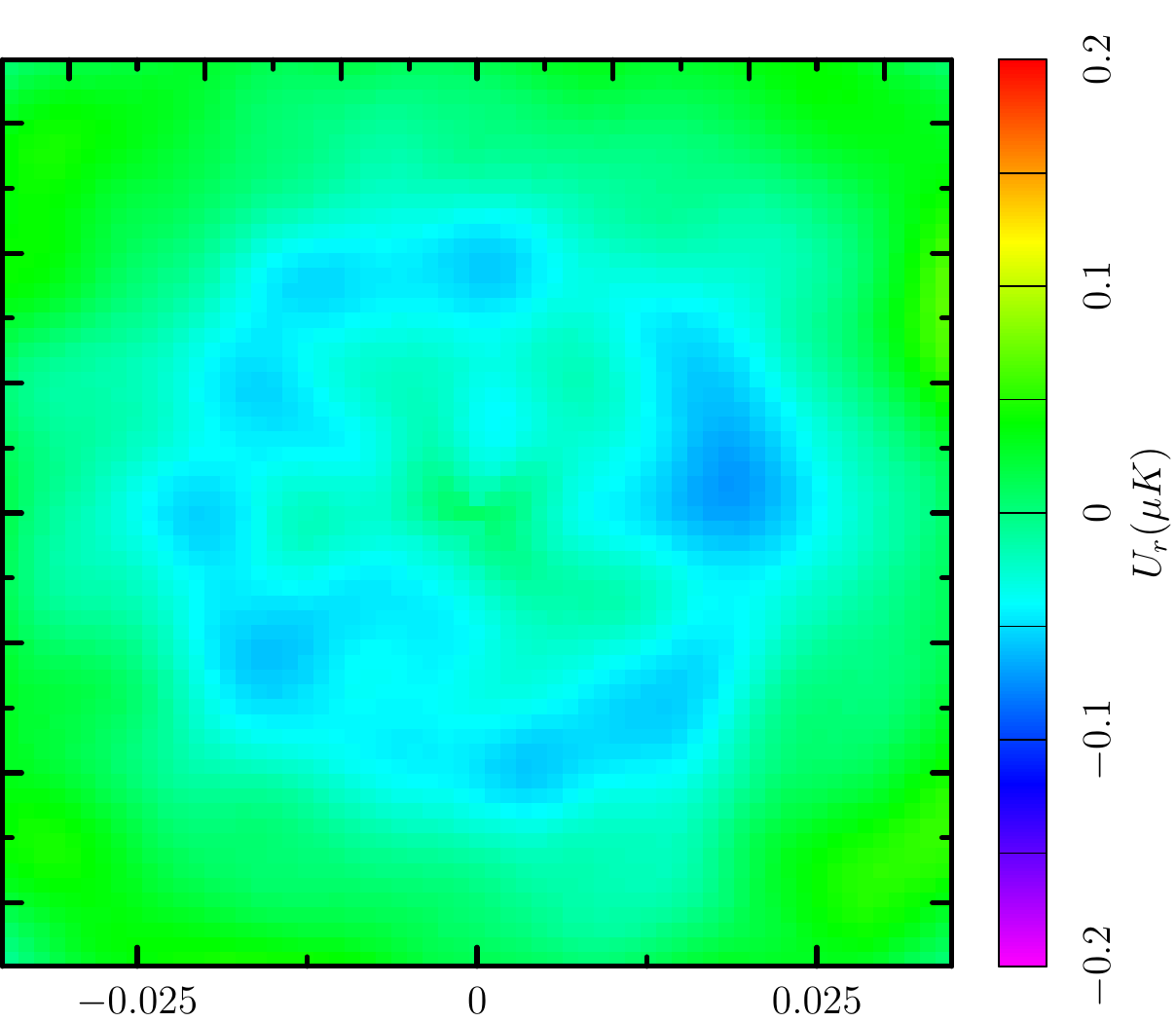}
  \caption{The stacked $[-2^\circ, 2^\circ] \times [-2^\circ, 2^\circ]$ ($[-0.035, 0.035] \times [-0.035, 0.035]$ in radians) patches of $Q_r$ (upper panels) and $U_r$ (lower panels) around the hottest pixels that cover $5\%\times 4\pi$ steradian in the LR63 mask. The colors represent the mean values and the headless vectors in the upper panels show the polarization directions. The left column shows Planck  353\,GHz half-mission 2 ($100\lesssim\ell\le 600$ band-passed) polarization maps stacked around the hottest pixels of Planck 353\,GHz half-mission 1 ($100\lesssim\ell \le 600$ band-passed) intensity map. The right column shows the same for a DUSTFILAMENTS simulation. \label{fig:stack}}
\end{figure*}

In Fig.~\ref{fig:stack} we show the stacked $Q_r$, $U_r$ patches around the hottest pixels that cover $5\%\times 4\pi $ steradian  within the LR63 mask. The left panels are the Planck result, and the right panels are done with the DUSTFILAMENTS simulation in Ref.~\cite{DUSTFILAMENTS}, which is publicly downloadable. The similar patterns of stacked $Q_r$  in the upper panels confirm that the simple filament-based approximation can reproduce the observed dust $TE$ correlation to a reasonably good accuracy. The rich ring pattern of stacked $U_r$ in Planck maps, however, is not seen in the DUSTFILAMENTS simulation. The much weaker $U_r$ signal in DUSTFILAMENTS simulation (faint blue ring in the lower right panel) arises from statistical fluctuations, or in cosmological terms, the cosmic variance of the particular simulation used here. 

The above example gives a good visual illustration that the typical $TB$ signal (stacked $U_r$) of DUSTFILAMENTS simulations is much smaller than the Planck observed one. The stacking approach, however, is not an optimal method for quantitative calculation of the statistics. This is because a large set of simulations is needed to compute the non-diagonal covariance matrix, and possibly non-Gaussian features of the stacked $U_r(\omega)$ vector. Fourier-space modes are less coupled and hence are better for a statistical analysis, which we carry out below.

\section{Statistical Analysis}

Since we are mostly interested in a coherent nonzero $TB$ signal over a wide range of angular scales ($40\lesssim \ell \lesssim 600$), a full likelihood analysis of $TB$ power spectrum, which is computationally expensive, may be avoided by compressing the information to an integrated quantity
\begin{equation}
  \xi_p \equiv \frac{\sum_{\ell = \ell_{\min}}^{\ell_{\max}} C_{\ell}^{TB}\ell^{-\alpha}}{\sum_{\ell = \ell_{\min}}^{\ell_{\max}} C_{\ell}^{TE}\ell^{-\alpha}}. \label{eq:xi}
\end{equation}
Unless otherwise stated, we fix the scale mask to be $[\ell_{\min}, \ell_{\max}]=[40, 600]$, the multipole range where the positive $TB$ signal was found in Planck 353\,GHz maps. The factor $\ell^{-\alpha}$, where $\alpha=2.44$ is roughly the spectral index of dust $TE$ and $TB$ power spectra found by Planck mission~\cite{PlanckXXX, Planck2018FG}, approximately sets equal weights  to all multipoles. The weighted sum of $C_\ell^{TE}$ in the denominator makes $\xi_p$ insensitive to the overall normalizations of the simulated temperature and polarization maps. Note that the power spectra and hence $\xi_p$ also depend on the sky mask, because the dust maps are not statistically isotropic. 

\begin{figure}
  \centering
  \includegraphics[width=0.5\textwidth]{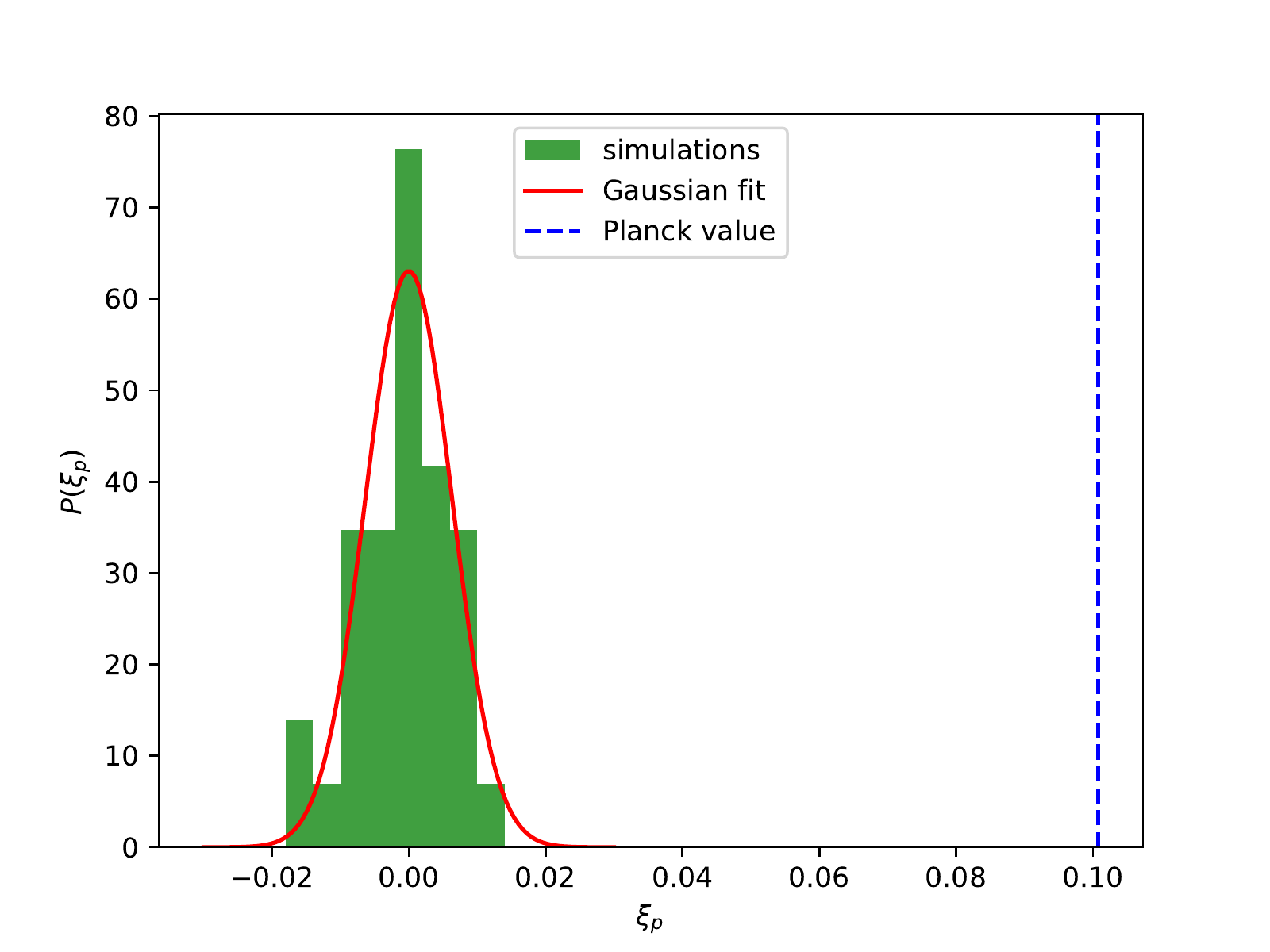}
  \caption{Gaussian fit to the histogram of LR63 masked $\xi_p$ from 36 DUSTFILAMENTS simulations. The Planck measured $\xi_p$ is in $15.9\sigma$ tension with the Gaussian fit.\label{fig:xi}}
\end{figure}

According to the central limit theorem, the integrated quantity $\xi_p$ should approximately obey a Gaussian distribution. This greatly simplify the problem, as we can then compute the probability density function $P(\xi_p)$ with a small set of simulations. Fig.~\ref{fig:xi} shows the result from 36 DUSTFILAMENTS simulations, with sky mask LR63 and scale mask $[\ell_{\min},\ell_{\max}] = [40, 600]$. The Planck $\xi_p$ is about $16\sigma$ away from the simulated distribution, suggesting that simulations with symmetric distribution of magnetic misalignment angles fails to describe the reality.

\begin{table*} 
  \caption{$\xi_p$ statistics from 36 DUSTFILAMENTS simulations. The last two columns show Planck $\xi_p$ value and its tension with the simulations.\label{tab:xi}}
  \begin{tabular}{cccccc}
    \hline
    \hline
    \textbf{scale mask $[\ell_{\min},\ell_{\max}]$} &  \textbf{sky mask}	& \textbf{standard deviation}	& \textbf{excess kurtosis}  & \textbf{Planck $\xi_p$} & \textbf{tension} \\
    \hline
    &  LR72		&  $0.0054$	& $-0.98$ & $0.068$ & $12.6\sigma$ \\
    &  LR63		& $0.0063$	& $-0.13$ & $0.101$ & $15.9\sigma$ \\
    $[40, 600]$ &  LR53		& $0.0072$	& $-0.33$ & $0.091$ & $12.6\sigma$ \\
    &  LR42		& $0.0080$	& $-0.94$ & $0.138$ & $17.3\sigma$ \\
    &  LR33		& $0.0084$	& $-0.43$ & $0.199$ & $23.8\sigma$ \\
    &  LR24		& $0.0096$	& $-0.48$ & $0.183$ & $19.2\sigma$\\
    \hline
    $[80, 300]$ & LR63 & $0.0114$ & $-0.13$   & $0.122$ & $10.6\sigma$\\
    \hline
  \end{tabular}
\end{table*}

\begin{figure}
  \centering
  \includegraphics[width=0.5\textwidth]{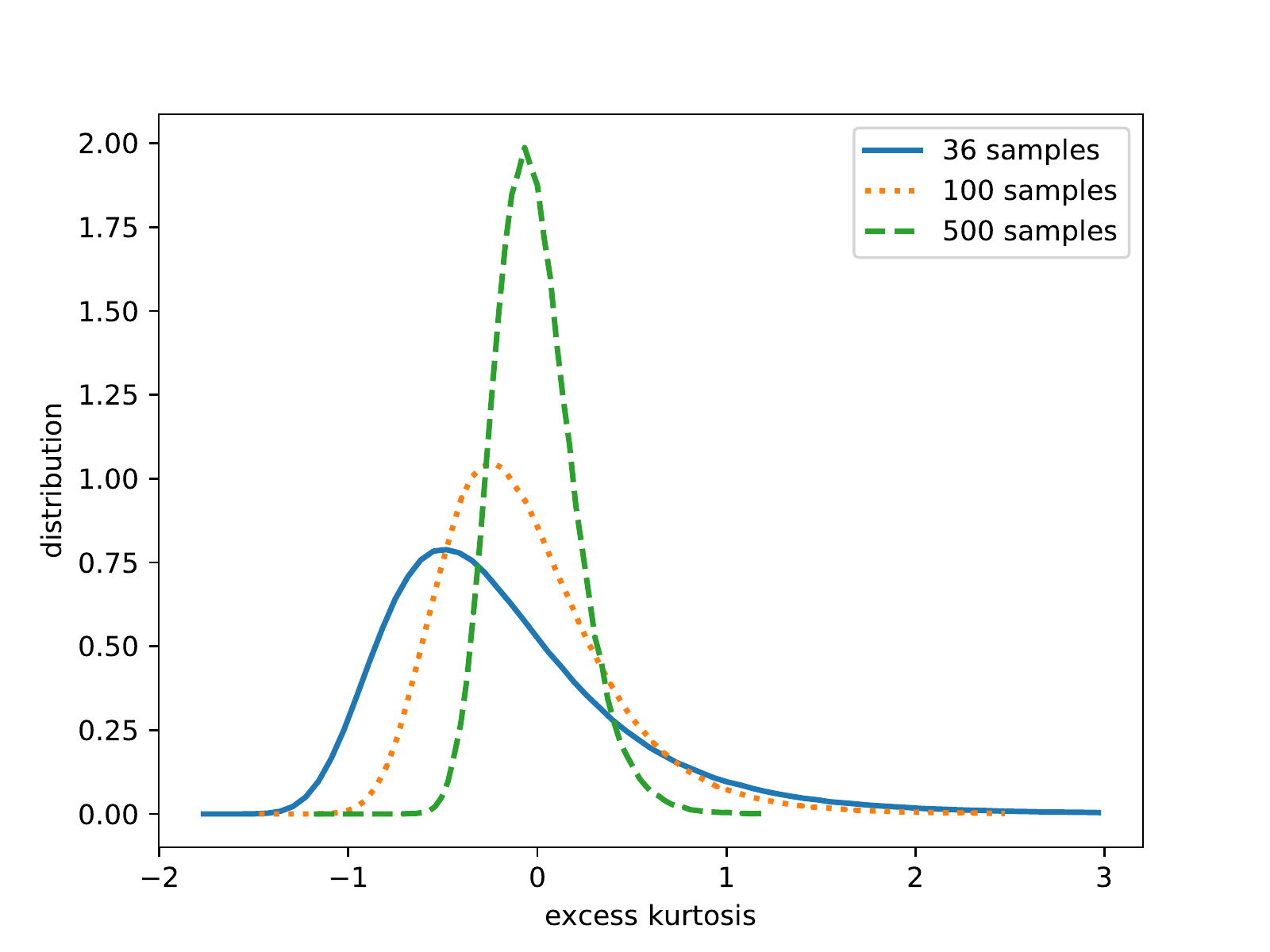}
  \caption{Distribution of excess kurtosis of a given number of random Gaussian samples. \label{fig:kurt}}
\end{figure}

To further test the look-elsewhere effect,  we show in Table~\ref{tab:xi} the results for various choices of sky mask and scale mask. In all cases, the tension between Planck data and simulations exceeds $10\sigma$, strongly rejecting the simulations as an acceptable description of the reality. We also list the excess kurtosis as a measure of the deviation from Gaussian tail distribution. For a comparison, in Fig.~\ref{fig:kurt} we show the distribution of excess kurtosis of 36 samples drawn from a perfect Gaussian distribution. The kurtosis values in Table~\ref{tab:xi} are all in good agreement with the distribution shown in Fig.~\ref{fig:kurt}, indicating no evidence of non-Gaussianity in $P(\xi_p)$.

Unlike the CMB statistics that is known to be  isotropic and very close to Gaussian, the statistics of Galactic dust emission is expected to be very non-Gaussian and spatially dependent. Although the central limit theorem guarantees that the integrated quantity $\xi_p$ approximately follows a Gaussian distribution, it is hard to quantify the level of proximity between $P(\xi_p)$ and a perfect Gaussian function. Ideally one would also like to numerically test the convergence of $P(\xi_p)$ towards a Gaussian distribution by increasing the number of simulations. As shown in Fig.~\ref{fig:kurt}, however, the distribution width of excess kurtosis does not shrink much if we increase the number of simulations from 36 to 100. Without investing much more computer power, we will not be able to tell whether the negative values of excess kurtosis in Table~\ref{tab:xi} are due to statistical fluctuations or are consequences of less extended tail distributions of $\xi_p$. Nevertheless, the latter case (less extended tails of $P(\xi_p)$) increases the tension between data and the null hypothesis, and therefore can only make our conclusion more robust.

\section{Conclusions}

The major conclusion of this work, that the observed dust $TB$ signal  is inconsistent with a pure statistical fluctuation, may sound a bit trivial in the sense that the cosmic variance (statistical fluctuations) of high-$\ell$  power spectra are all suppressed. However, as one can see from the lower right panel of Fig.~\ref{fig:stack}, even the input model perfectly preserves parity symmetry, a particular realization may still present recognizable $TB$ patterns at sub-degree scales. Thus, a quantitative estimation of the tail distribution of $TB$ correlation is still necessary, especially when the observed dust $TB$ signal is small and the dust maps are highly non-Gaussian.

Since the null hypothesis of no preferred alignment handedness is ruled out,  at least around a large neighborhood ($\sim$ a few hundred $\mathrm{pc}$) of the solar system, there must exist a globally preferred handedness of the filament magnetic misalignment. Recent studies on the correlation between Farady depth and polarization of the synchrotron emission suggest helicity in the large-scale Galactic magnetic field~\cite{West20}, which may lead to a dust $TB$ correlation~\cite{Bracco19}. However, this explanation has only been shown to be applicable to $TB$ correlation on large angular scales ($\ell \lesssim 50$)~\cite{Bracco19}. The physical origin of the parity violation on small scales (up to $\ell\sim 600$) remains unknown. Understanding the physical origin of this parity violation potentially has a great value for many fields of astrophysics and cosmology. We also warn that the parity violation of the dust foreground must be carefully taken into account when studying the recently discovered tantalizing $\sim 2$-$3\sigma$ hints of cosmic birefringence~\cite{Eskilt22, Greco22, Diego22, Komatsu22}.

\section{Acknowledgements}
This work is supported by the National key R\&D Program of China (Grant No. 2020YFC2201600), National Natural Science Foundation of China (NSFC) under Grant No. 12073088, Guangdong Major Project of Basic and Applied Basic Research (Grant No. 2019B030302001), and National SKA Program of China No. 2020SKA0110402.


\begin{thebibliography}{10}

\bibitem{Planck2018Params}
N.~{Aghanim}, Y.~{Akrami}, M.~{Ashdown}, et~al.
\newblock {Planck 2018 results. VI. Cosmological parameters}.
\newblock {\em Astron. Astrophys.}, 641:A6, 2020.

\bibitem{Planck2018Inflation}
Y.~{Akrami}, F.~{Arroja}, M.~{Ashdown}, et~al.
\newblock {Planck 2018 results. X. Constraints on inflation}.
\newblock {\em Astron. Astrophys.}, 641:A10, 2020.

\bibitem{ACT20}
Simone {Aiola}, Erminia {Calabrese}, Lo{\"\i}c {Maurin}, et~al.
\newblock {The Atacama Cosmology Telescope: DR4 Maps and Cosmological
  Parameters}.
\newblock {\em JCAP}, 12:047, 2020.

\bibitem{SPT21Params}
L.~{Balkenhol}, D.~{Dutcher}, P.~A.~R. {Ade}, et~al.
\newblock {Constraints on \ensuremath{\Lambda}CDM extensions from the SPT-3G
  2018 EE and TE power spectra}.
\newblock {\em Phys. Rev. D}, 104(8):083509, 2021.

\bibitem{Polarbear20}
S.~{Adachi}, M.~A.~O. {Aguilar Fa{\'u}ndez}, K.~{Arnold}, et~al.
\newblock {A Measurement of the CMB $E$-mode Angular Power Spectrum at
  Subdegree Scales from670 Square Degrees of POLARBEAR Data}.
\newblock {\em Astrophys. J.}, 904(1):65, 2020.

\bibitem{BK16}
P.~A.~R. {Ade}, Z.~{Ahmed}, R.~W. {Aikin}, et~al.
\newblock {Improved Constraints on Cosmology and Foregrounds from BICEP2 and
  Keck Array Cosmic Microwave Background Data with Inclusion of 95 GHz Band}.
\newblock {\em Phys. Rev. Lett.}, 116:031302, 2016.

\bibitem{BK18}
P.~A.~R. {Ade}, Z.~{Ahmed}, R.~W. {Aikin}, et~al.
\newblock {BICEP2 / Keck Array x: Constraints on Primordial Gravitational Waves
  using Planck, WMAP, and New BICEP2/Keck Observations through the 2015
  Season}.
\newblock {\em Phys. Rev. Lett.}, 121:221301, 2018.

\bibitem{BK21}
M.~{Tristram}, A.~J. {Banday}, K.~M. {G{\'o}rski}, et~al.
\newblock {Improved limits on the tensor-to-scalar ratio using BICEP and Planck
  data}.
\newblock {\em Phys. Rev. D}, 105(8):083524, 2022.

\bibitem{PlanckXXX}
R.~{Adam}, P.~A.~R. {Ade}, N.~{Aghanim}, et~al.
\newblock {Planck intermediate results. XXX. The angular power spectrum of
  polarized dust emission at intermediate and high Galactic latitudes}.
\newblock {\em \aap}, 586:A133, February 2016.

\bibitem{Planck2018FG}
Y.~{Akrami}, M.~{Ashdown}, J.~{Aumont}, et~al.
\newblock {Planck 2018 results. XI. Polarized dust foregrounds}.
\newblock {\em Astron. Astrophys.}, 641:A11, 2020.

\bibitem{BK15}
P.~A.~R. {Ade}, N.~{Aghanim}, Z.~{Ahmed}, R.~W. {Aikin}, et~al.
\newblock {Joint Analysis of BICEP2/$Keck Array$ and $Planck$ Data}.
\newblock {\em Phys. Rev. Lett.}, 114:101301, 2015.

\bibitem{Purcell75}
E.~M. {Purcell}.
\newblock {\em The Dusty Universe}.
\newblock 1975.

\bibitem{DUSTGrain}
B.~T. {Draine}.
\newblock {Interstellar Dust Grains}.
\newblock {\em Annual Review of Astronomy \& Astrophysics}, 41:241--289,
  January 2003.

\bibitem{Bracco19}
A.~{Bracco}, S.~{Candelaresi}, F.~{Del Sordo}, and A.~{Brandenburg}.
\newblock {Is there a left-handed magnetic field in the solar neighborhood?.
  Exploring helical magnetic fields in the interstellar medium through dust
  polarization power spectra}.
\newblock {\em \aap}, 621:A97, January 2019.

\bibitem{Eskilt22}
J.~R. {Eskilt}.
\newblock {Frequency-dependent constraints on cosmic birefringence from the LFI
  and HFI Planck Data Release 4}.
\newblock {\em \aap}, 662:A10, June 2022.

\bibitem{Greco22}
Alessandro {Greco}, Nicola {Bartolo}, and Alessandro {Gruppuso}.
\newblock {Cosmic birefrigence: cross-spectra and cross-bispectra with CMB
  anisotropies}.
\newblock {\em \jcap}, 2022(3):050, March 2022.

\bibitem{Diego22}
P.~{Diego-Palazuelos}, J.~R. {Eskilt}, Y.~{Minami}, M.~{Tristram}, R.~M.
  {Sullivan}, A.~J. {Banday}, R.~B. {Barreiro}, H.~K. {Eriksen}, K.~M.
  {G{\'o}rski}, R.~{Keskitalo}, E.~{Komatsu}, E.~{Mart{\'\i}nez-Gonz{\'a}lez},
  D.~{Scott}, P.~{Vielva}, and I.~K. {Wehus}.
\newblock {Cosmic Birefringence from Planck Public Release 4}.
\newblock {\em arXiv e-prints}, page arXiv:2203.04830, March 2022.

\bibitem{Komatsu22}
Eiichiro {Komatsu}.
\newblock {New physics from the polarised light of the cosmic microwave
  background}.
\newblock {\em arXiv e-prints}, page arXiv:2202.13919, February 2022.

\bibitem{Weiland20}
J.~L. {Weiland}, G.~E. {Addison}, C.~L. {Bennett}, M.~{Halpern}, and
  G.~{Hinshaw}.
\newblock {An Examination of Galactic Polarization with Application to the
  Planck TB Correlation}.
\newblock {\em \apj}, 893(2):119, April 2020.

\bibitem{MHD1}
Chang-Goo {Kim}, Steve~K. {Choi}, and Raphael {Flauger}.
\newblock {Dust Polarization Maps from TIGRESS: E/B Power Asymmetry and TE
  Correlation}.
\newblock {\em \apj}, 880(2):106, August 2019.

\bibitem{MHD2}
A.~G. {Kritsuk}, S.~D. {Ustyugov}, and M.~L. {Norman}.
\newblock {The structure and statistics of interstellar turbulence}.
\newblock {\em New Journal of Physics}, 19(6):065003, June 2017.

\bibitem{MHD3}
Robert~R. {Caldwell}, Chris {Hirata}, and Marc {Kamionkowski}.
\newblock {Dust-polarization Maps and Interstellar Turbulence}.
\newblock {\em \apj}, 839(2):91, April 2017.

\bibitem{PlanckEBAsym}
P.~A.~R. {Ade}, N.~{Aghanim}, M.~{Arnaud}, et~al.
\newblock {Planck intermediate results. XXXVIII. E- and B-modes of dust
  polarization from the magnetized filamentary structure of the interstellar
  medium}.
\newblock {\em Astron. Astrophys.}, 586:A141, 2016.

\bibitem{DUSTFILAMENTS}
Carlos {Herv{\'\i}as-Caimapo} and Kevin~M. {Huffenberger}.
\newblock {Full-sky, Arcminute-scale, 3D Models of Galactic Microwave
  Foreground Dust Emission Based on Filaments}.
\newblock {\em \apj}, 928(1):65, March 2022.

\bibitem{Huffenberger20}
Kevin~M. {Huffenberger}, Aditya {Rotti}, and David~C. {Collins}.
\newblock {The Power Spectra of Polarized, Dusty Filaments}.
\newblock {\em \apj}, 899(1):31, August 2020.

\bibitem{Clark21}
S.~E. {Clark}, Chang-Goo {Kim}, J.~Colin {Hill}, and Brandon~S. {Hensley}.
\newblock {The Origin of Parity Violation in Polarized Dust Emission and
  Implications for Cosmic Birefringence}.
\newblock {\em \apj}, 919(1):53, September 2021.

\bibitem{Planck2018HFI}
N.~{Aghanim}, Y.~{Akrami}, M.~{Ashdown}, et~al.
\newblock {Planck 2018 results. III. High Frequency Instrument data processing
  and frequency maps}.
\newblock {\em \aap}, 641:A3, September 2020.

\bibitem{SRoll2}
J.~M. {Delouis}, L.~{Pagano}, S.~{Mottet}, J.~L. {Puget}, and L.~{Vibert}.
\newblock {SRoll2: an improved mapmaking approach to reduce large-scale
  systematic effects in the Planck High Frequency Instrument legacy maps}.
\newblock {\em \aap}, 629:A38, September 2019.

\bibitem{Konstantinou22}
A.~{Konstantinou}, V.~{Pelgrims}, F.~{Fuchs}, and K.~{Tassis}.
\newblock {Polarization power spectra and dust cloud morphology}.
\newblock {\em arXiv e-prints}, page arXiv:2204.13127, April 2022.

\bibitem{Healpix}
K.~M. {G{\'o}rski}, E.~{Hivon}, A.~J. {Banday}, B.~D. {Wandelt}, F.~K.
  {Hansen}, M.~{Reinecke}, and M.~{Bartelmann}.
\newblock {HEALPix: A Framework for High-Resolution Discretization and Fast
  Analysis of Data Distributed on the Sphere}.
\newblock {\em \apj}, 622(2):759--771, April 2005.

\bibitem{NaMaster}
David {Alonso}, Javier {Sanchez}, An{\v{z}}e {Slosar}, and {LSST Dark Energy
  Science Collaboration}.
\newblock {A unified pseudo-C$_{{\ensuremath{\ell}}}$ framework}.
\newblock {\em \mnras}, 484(3):4127--4151, April 2019.

\bibitem{WMAP7}
E.~{Komatsu}, K.~M. {Smith}, J.~{Dunkley}, et~al.
\newblock {Seven-year Wilkinson Microwave Anisotropy Probe (WMAP) Observations:
  Cosmological Interpretation}.
\newblock {\em \apjs}, 192(2):18, February 2011.

\bibitem{Planck2016IandS}
P.~A.~R. {Ade}, N.~{Aghanim}, Y.~{Akrami}, et~al.
\newblock {Planck 2015 results. XVI. Isotropy and statistics of the CMB}.
\newblock {\em \aap}, 594:A16, September 2016.

\bibitem{Note1}
For the $4^\circ \times 4^\circ $ stacking in the present work, the difference
  between $\theta $ and $2\protect \qopname \relax o{sin}\protect \frac {\theta
  }{2}$ is actually unimportant.

\bibitem{West20}
J.~L. {West}, R.~N. {Henriksen}, K.~{Ferri{\`e}re}, A.~{Woodfinden},
  T.~{Jaffe}, B.~M. {Gaensler}, and J.~A. {Irwin}.
\newblock {Helicity in the large-scale Galactic magnetic field}.
\newblock {\em \mnras}, 499(3):3673--3689, December 2020.

\end{thebibliography}

\end{document}